\documentclass[twocolumn,prb,showpacs,prbprintnumbers,fleqn,superscriptaddress]{revtex4}
\usepackage{epsfig}
\usepackage{graphicx}
\usepackage{amsfonts}

\newcommand{\ct}{\cite}
\newcommand{\bi}{\bibitem}

\newcommand{\bra}{\langle}
\newcommand{\ket}{\rangle}
\newcommand{\non}{\nonumber}

\newcommand{\be}{\begin{equation}}
\newcommand{\ee}{\end{equation}}
\newcommand{\ba}{\begin{eqnarray}}
\newcommand{\ea}{\end{eqnarray}}

\begin{document}
\title{ Effects of interference in the dynamics of spin-1/2 transverse XY Chain driven periodically through quantum critical points}
\author{Victor Mukherjee}
\email{victor@iitk.ac.in}
\author{Amit Dutta}
\email{dutta@iitk.ac.in}
\affiliation{Department of Physics, Indian Institute of Technology, Kanpur 
208 016, India}

\date{\today}

\begin{abstract}

We study the effects of interference on the quenching dynamics of a one-dimensional spin $1/2$ $XY$ model in the presence of a transverse field ($h(t)$) which varies sinusoidally with time as $h = h_0\cos{\omega t}$, with $|t| \leq t_f = \pi/\omega$. We have explicitly shown that the finite values of $t_f$ make the dynamics inherently dependent on the phases of probability amplitudes, which had been hitherto unseen in all cases of linear quenching with large initial and final times. In contrast, we also consider the situation where the magnetic field consists of an oscillatory as well as a linearly varying component, i.e.,   $h(t) = h_0\cos{\omega t} + t/\tau$, where the interference effects lose importance in the limit of large $\tau$. Our purpose is to estimate the defect density and the local entropy density in the final state if the
system is initially prepared in its ground state.  For a single crossing through the quantum critical point with $h= h_0\cos{\omega t}$, the density of defects in the final state is calculated by mapping the dynamics to an equivalent Landau Zener problem by linearizing near the crossing point, and is found to vary as $\sqrt{\omega}$ in the limit of small $\omega$. On the other hand, the local entropy density is found to attain a maximum as a function of $\omega$  near a characteristic scale $\omega_0$. Extending to the situation of multiple  crossings, we show that the role of finite initial and final times of quenching are manifested non-trivially in the  interference effects  of certain resonance modes which solely contribute to  the production of defects. Kink density as well as the diagonal entropy density show oscillatory dependence on the number of full cycles of oscillation. Finally, the inclusion of a linear term in the transverse field on top of the oscillatory compo
 nent, results to a kink density which decreases continuously with $\tau$ while increases monotonically with $\omega$. The entropy density also shows monotonous change with the parameters, increasing with $\tau$ and decreasing with $\omega$,  in sharp contrast to the situations studied  earlier. We do also propose appropriate scaling relations for the defect density in above situations and compare the
results with the numerical results obtained by integrating the Schr\"{o}dinger equations.

\end{abstract}

\maketitle

\section{Introduction}

The phenomena of quantum phase transitions occurring at absolute zero temperature have attracted serious attention of scientists
in recent years \ct{sachdev99,dutta96}. A quantum critical point is associated with  a diverging length scale ($\xi$)  and a diverging time scale  ($\xi_{\tau}$) which satisfy the
scaling forms $\xi \sim |\overline d|^{-\nu}$ and $\xi_{\tau} \sim {\xi}^{z}$ in the vicinity of the quantum critical point. Here $\overline d$ denotes the deviation from the critical point, and $\nu$, $z$ are the corresponding critical exponents. Following
the possibility of experimental studies of quantum systems trapped in optical lattices \ct{bloch08},
there is a recent interest in theoretical studies of the related models \ct{sengupta04,bose}. When a parameter of the Hamiltonian of the system is swept across  a quantum critical point.
The diverging relaxation time near the quantum critical point forces the system to undergo non-adiabatic evolution irrespective of the rate of change of parameters.
If the system  is in its ground state at initial time, non-adiabatic transitions  are manifested in the occurrence of non-zero 'defects' (called kinks) and non-zero local entropy in the system. 

According to the Kibble-Zurek argument, if a parameter of the Hamiltonian is varied as $t/\tau$, the density of defects ($n$) in the final state is expected to scale as $n \sim \tau^{-\nu d / (\nu z + 1)}$, where $d$ is the spatial dimensionality of the system
\ct{kibble76,zurek96,zurek05,polkovnikov05}.The above scaling form has been verified for quantum
spin systems quenched through critical points \ct{dziarmaga05, levitov06, mukherjee07, divakaran07} and also
generalized to the cases of non-linear quenching\ct{sen08} when a parameter is quenched as $h(t)\sim |t/\tau|^{|a|}sgn(t)$, for gapless systems \ct{sengupta08} and also for quantum systems with disorder \ct{dziarmaga06} and systems coupled to external environment \ct{patane08}. Recently, a generalized form of the Kibble-Zurek scaling has been introduced which includes a situation where the system is quenched through the multicritical point\ct{divakaran08} which shows that the general expression for kink density can be given as $n \sim {\tau}^{-d/2z_2}$, where $z_2$ determines the scaling of
the off-diagonal term of the equivalent Landau-Zener problem close to the critical point.

In parallel to the studies on spin chains in condensed matter physics, great progress has also been made in the realm of quantum optics in exploring the dynamics of two level systems undergoing landau Zener tunneling due to oscillatory temporal variation of the parameters \ct{kayanuma92, kayanuma94, ashhab07, hanggi05}. The Landau Zener transition probabilities have been calculated for single crossing as well as for multiple crossings. Superposition of a linear field along with the sinusoidally varying energy levels gives rise to an altogether different situation, which has also been studied thoroughly in recent years \ct{kayanuma00}.

In our present work we exploit the techniques used in the above papers to explore the dynamics of one dimensional spin 1/2 chain undergoing quantum phase transitions due to the application of an oscillatory or an oscillatory as well as a linearly varying magnetic field and compare the results with the earlier findings. Investigation of the dynamics of one dimensional transverse XY model under repeated quenching of linearly varying transverse magnetic field has been carried out in a recent work \ct{mukherjee08} and was shown that the defect density decreases in the reverse path though the entropy density increases monotonically with the number of quenching. However, the scaling of the defect density and the local entropy density when the quantum critical point is crossed due to a sinusoidal variation of the magnetic field has not been attempted before. Sinusoidal quenching puts an upper bound to the initial and final times, which makes the process of coarse graining invalid in
  the present scenario. The resultant dynamics of the system becomes dependent on the phases of the probability amplitudes, leading to the occurance of constructive or destructive interferences.  The spin-1/2 transverse XY chain \ct{lieb61} discussed in this paper is described by the Hamiltonian
\be H ~=~ - \frac{1}{2} ~\sum_n ~(J_x \sigma^x_n \sigma^x_{n+1} + J_y \sigma^y_n \sigma^y_{n+1} + h \sigma^z_n), \label{h1} \ee 
where the $\sigma$'s are Pauli spin matrices satisfying the usual 
commutation relations and the  interactions and the transverse field are denoted by 
 $J_x, J_y$ and $h$, respectively, with  $J_x$, $J_y$ and $h$  all non-random.  The interaction strength $J_x$ and $J_y$  are always time-independent whereas we shall
use time-dependent transverse field in the subsequent sections.

The Hamiltonian in Eq. (\ref{h1}) can be exactly diagonalized using the 
Jordan-Wigner (JW) transformation which maps a system of spin-1/2's to a system 
of spinless fermions \ct{lieb61,bunder99} given by
\ba c_n &=& \exp \left( i \pi \sum_{j=1}^{n-1} \sigma_j^z \right) \sigma_n^- ,
 \ea
where $\sigma_n^{\pm} = (\sigma_n^x \pm i \sigma_n^y)/2$  and the operator $\sigma_n^z = 2 c_n^{\dagger}c_n - 1$ and the operators $c_n$ satisfy the fermionic anti-commutation
relations. In terms of JW fermions, the above Hamiltonian can be rewritten in Fourier space with a periodic boundary condition as
\ba H &=& - ~\sum_{k\rangle0} ~\{ ~[(J_x + J_y) \cos k +h] ~(c_k^{\dagger} c_k +
c_{-k}^{\dagger} c_{-k}) \non \\
& & ~~~~~~~~~~~~~~+ i (J_x -J_y) \sin k ~(c_k^{\dagger} c_{-k}^{\dagger}
+ c_k c_{-k}) \} . \label{h2} \ea
Diagonalizing the Hamiltonian in terms of the Bogoliubov fermions, we arrive 
at an expression for the gap in the excitation spectrum given by 
\ct{lieb61,bunder99}
\be \epsilon_k = [h^2 +J_x^2 + J_y^2 + 2 h (J_x + J_y) \cos k + 2 J_x J_y
\cos 2k ]^{1/2}. \label{ek} \ee
The gap given in Eq. (\ref{ek}) vanishes at $h = \mp (J_x + J_y)$ for wave 
vectors $k =0$ and $\pi$ respectively, signaling a quantum phase transition 
from a ferromagnetically ordered phase to a quantum paramagnetic phase known 
as the ``Ising" 
transition \ct{bunder99}. On the other hand, the vanishing of gap at  $J_x =J_y$ for $|h|<(J_x+J_y)$
at an ordering wave-vector $k_0 = \cos^{-1}(-h/2J_x)$ signifies a quantum phase
transition, belonging to a different universality class from the Ising transitions, between two ferro-magnetically ordered phases. 

The advantage of employing the JW transformation is that 
when projected to the two-dimensional subspace spanned by $|0\ket$ and 
$|{k,-k}\ket$, the Hamiltonian takes the form $H = \sum_{k} H_k$ where
the reduced Hamiltonian is written in the form
\ba H_k (t) &=& -~[ h ~+~ (J_x+J_y)\cos k] ~I_2 \non \\
&+& \left[ \begin{array}{cc} h+ (J_x + J_y) \cos k & i (J_x - J_y)
\sin k \\
-i(J_x - J_y) \sin k & -h-(J_x+J_y)\cos k \end{array} \right], \non \ea
where $I_2$ denotes the $2 \times 2$ identity matrix. Therefore, the many-body
problem is effectively reduced to the problem of a two-level 
system with two levels denoted by the states $|0\ket$ and $|{k,-k}\ket $ which we 
refer to as diabatic basis vectors. We shall denote the basis vectors as $|0\ket$ and $|k,-k\ket$ as $|1 \ket$ and $|2\ket$,
respectively in this work for notational convenience and refer the states as diabatic
energy levels.

The paper is organized in the following way: We have already discussed the model we are going to study.
In section II, we discuss the case of oscillatory magnetic field but restrict our attention to the
situation when the quantum critical points are crossed only once while in section III the possibility
of multiple crossing is included. Section IV is used to discuss the dynamics with a magnetic field
which has both linearly varying and oscillatory components. In every situation results
obtained through approximate analytical methods are contrasted with the numerical ones
obtained by direct integration. Conclusion and summary of the results
are presented in the last section.

\section{Oscillatory quenching through a quantum critical point: single crossing}

In this section, we shall study the spin chain driven by an oscillatory
transverse field given by $h(t)= h_0 \cos \omega t$ from an initial time $-\pi/\omega$
to a final time $0$ so that it crosses the gapless point only once during
the course of evolution. The system initially prepared in the ground state $|1 \ket$
whereas  the final ground state is $|2 \ket$. We shall evaluate the probability of the
state  $|1 \ket$ in the final state due to non-adiabatic evolution through the gapless
point.  As discussed above,  the transverse XY chain Hamiltonian can be written as direct sum of  $2 \times 2$
reduced Hamiltonian matrices $H_k$.  For an oscillatory  transverse field, the
reduced  Hamiltonian gets modified to

\ba &&H_k (t) = \non \\  && \left[ \begin{array}{cc} h_0 \cos {\omega t} + (J_x + J_y) \cos k & i (J_x - J_y)\sin k \\
-i(J_x - J_y) \sin k & -h_0 \cos {\omega t}-(J_x+J_y)\cos k \end{array} \right], \non \ea
For any given mode $k$, the instantaneous energy gap of the Hamiltonian is minimum
at a time $t_{0,k}$ such that $h_0 \cos {\omega t_{0,k}} + (J_x + J_y) \cos k=0$ where the
diabatic levels cross each other. On the other
hand, the energy gap vanishes for the wavevectors $k=0$ and $k=\pi$ at a time $\cos \omega t=
\mp (J_x +J_y)\cos{k}/h_0$, respectively, signalling the quantum phase transition of
Ising class.

Denoting the probability amplitudes for the states $|1\ket$ and $|2\ket$ as $\overline {C_{1,k}}(t), \overline {C_{2,k}}(t)$, respectively, a general state vector in the reduced Hilbert space is witten as.
\be
\psi_k(t) = \overline {C_{1,k}}(t)|1\ket + \overline {C_{2,k}}(t)|2\ket
\ee
Henceforth, we set  $(J_x + J_y) = J$, and modulus of the off-diagonal terms = $|J_x - J_y|\sin k$ is denoted by $\Delta_k$. Also the modulus of rate of change of the diagonal terms at time $t=t_{0,k}$ (given by $h_0\omega \sin {\omega t_{0,k}}$) is denoted bt $\alpha_k$. Using the transformations 
\ba
&&\overline {C_{1,k}}(t) = C_{1,k}(t)e^{-i\int^{t} {dt^{'} [h_0 \cos {\omega t} + J \cos k ]}} \non \\
&&\overline {C_{2,k}}(t) = C_{2,k}(t)e^{-i\int^{t} {dt^{'} [-h_0 \cos {\omega t} - J \cos k ]}}
\ea
we can rewrite the Schr\"odinger equation describing the time evolution of the probability amplitudes in the  form
\ba
i\frac{d C_{1,k}(t)}{dt} &=&  \Delta_k C_{2,k}(t) e^{2i\int^{t}{dt^{'}[h_0 \cos {\omega t} + J \cos k ]}} \non \\
i\frac{d C_{2,k}(t)}{dt} &=&  \Delta_k C_{1,k}(t) e^{-2i\int^{t}{dt^{'}[h_0 \cos {\omega t} +  \cos k ]}}.
\ea
It should be noted that for large values of $[h_0 \cos {\omega t} + J \cos k ]$, the phase factors on the r.h.s. of eq. (7) oscillate rapidly in time.  As a result  the amplitudes $C_{1,k}(t)$, $C_{2,k}(t)$, averaged over small intervals of time, remain basically constant in time. On the other hand, close to $t = t_{0k}$, the phase factors assume stationary values, thus leading to non-adiabatic transition between the energy levels \ct{landau, suzuki05, hanggi05}. 

In this section, we prepare  the system in the ground state  with  initial conditions at $\omega t = -\pi$, i.e., $C_{1,k}(-\pi) = 1$ and $C_{2,k}(-\pi) = 0$ and the state evolves to $t=0$ so that the spin chain crosses the gapless quantum critical point only once. Using Eqs.~(7), one can arrive at the differential equation describing the amplitude $C_{1,k}(t)$ given by
\be
\frac{d^2 C_{1,k}}{d^2 t} - 2i (h_0 \cos \omega t+J\cos k) \frac {d C_{1,k}}{dt} + \Delta_k^2 C_{1,k} = 0
\ee
with the probability of defect in the final state at $t = 0$ given by $p_k =|C_{1,k}(0)|^2$.
The maximum contribution to the non-adiabatic transition probability comes from near the points where the energy gap between the
instantaneous levels is minimum. We can therefore linearize the term $h_0 \cos \omega t$ in the neighbourhood of $t_{0,k}$
to get
\be
\frac{d^2 C_{1,k}}{d^2 t} - 2i (-h_0 \omega (t-t_{0,k})\sin \omega t_{0,k} ) \frac {d C_{1,k}}{dt} + \Delta_k^2 C_{1,k} = 0
\ee
Eq.~(9) resembles the standard Landau Zener transition problem \ct{suzuki05} for the linear
quenching of the magnetic field with a variation
of the field $h=t/\tau_{\rm eff}$ where $\tau_{\rm eff}$ is given by the  
the rate of change of the diagonal terms of the Hamiltonian (1). Therefore, let us define $\alpha_k =  1/\tau_{\rm eff} =\frac{d}{dt} (\epsilon_1 - \epsilon_2)|_{t_{0,k}}= 2h_0 \omega \sin {\omega t_{0,k}} = 2\omega \sqrt {h^2_0 - J^2 \cos ^2_k}$.  The non-adiabatic excitation probability\ct{suzuki05} is given as $p_k = |C_{1,k}(0)|^2 = e^{-2\pi \gamma_k}$ , where $\gamma_k = \Delta ^2_k/|\frac{d}{dt} (\epsilon_1 - \epsilon_2)|_{t_{0,k}}$, leading
to
\ba 
p_k = e^{-\frac{\pi \Delta^2_k }{\omega \sqrt {h^2_0 - J^2\cos ^2 {k}}}}
\ea
At this point, the natural question to ask is that for what values of the parameter $h_0$ and $\omega$, the above relation of $p_k$ is
applicable. Of course, we have used the non-adiabatic transition probability of the standard
linear Landau Zener problem where time $t$ evolves from $-\infty$ to $+\infty$. The linearization near the crossing point employed above holds good only for small $\omega$. 
 More precisely, as discussed  below the linearization approximation is applicable when the 
the time period of one cycle of the magnetic field ($ 2\pi/\omega$) is much greater than the Landau Zener transition time ($T_{LZ,k}$) for 
a single crossing. The dimensionless Landau-Zener transition time \ct{vitanov99,vitanov96}  is defined as $\kappa_{LZ,k} = \sqrt{\alpha_k} T_{LZ,k} = |C_{2,k}(\overline t)|^2/\frac{d}{d\kappa}|C_{2,k}(0)|^2\approx {|C_{2,k}(+\infty)|^2}/(\frac{d}{d\kappa}|C_{2,k}(0)|^2)$, where $\kappa = \sqrt{\alpha_k}t$. Using the above definition, we find
 that $T_{LZ,k} \sim \Delta_k/\alpha_k $ in the adiabatic limit ($\Delta_k ^2/\alpha_k >> 1$) whereas in the non-adiabatic limit  (${\Delta ^2 _k}/{ \alpha_k} << 1$), $T_{LZ,k}$ is given as $T_{LZ,k} \sim {1}/{\sqrt{\alpha_k}}$.  It should be noted that for the linear quenching of the
magnetic field $h(t)=t/\tau$, $\alpha_k = 1/\tau$.

Generalizing to the case of periodic quenching,  $T_{LZ,k} \sim {\Delta_k}/({2\omega\sqrt{h^2_0 - J^2\cos^2 {k})}}$, in the adiabatic limit while in the non-adiabatic limit, $T_{LZ,k} \sim {1}/({\sqrt{2\omega\sqrt{h^2_0 - J^2\cos^2 {k}}}})$. The transitions are localized around $t_{0,k}$ as compared to the time period for one cycle of the magnetic field if $T_{LZ,k}$ is less than the time period of one oscillation, i.e., $T_{LZ,k} << \pi /\omega$. This means that  for $h^2_0 >> J^2$, in the adiabatic limit,  $\Delta_k << h_0$, and in the non-adiabatic limit, $\omega << h_0$.
 This leads to the conclusion that the equation (10) 
describing the non-adiabatic transition probability is valid only for large $h_0$ and small $\omega$. In the defect
density for small $\omega$ (i.e. $\omega < \pi (J_x - J_y)^2/\sqrt {h^2_0 - J^2}$)  only the modes close $k \sim 0, \pi$ contribute, resulting to a  kink density at
$t=0$ given by 
\ba
n \approx \frac{\sqrt {\omega \sqrt{h^2_0 - J^2}}}{\pi|J_x - J_y|} ~~~~~~~~~~ [\omega < \frac {\pi (J_x - J_y)^2}{\sqrt {h^2_0 - J^2}}] \ea

The analytical results for $p_k$ and hence the density of defects (obtained from Eq.~11) match exactly with the transition
probabilities obtained by numerical integration of the Schr\"{o}dinger Eqns for $h_0 >> \Delta_k, \omega $  
as shown in Figs.~ 1 and 2. From Eq.~11, we find that in the limit $h_0 >> J$,
the defect density shows a scaling form $n\sim (h_0 \omega)^{1/2}$ which can be generalized
using the Kibble-Zurek argument that assumes that the non-adiabatic transition is only
dominant at a time when the characteristic time scale of the system is of the order
of the rate of change of the Hamiltonian \ct{kibble76,zurek96,polkovnikov05}. In the limit of 
large $h_0$ and small $\omega$,
we can generalize the above prescription to derive a scaling form for the defect density
for a single crossing the quantum critical point due to the periodic driving of
the transverse field given by $n \sim (h_0 \omega)^{\nu d/(\nu z +1)}$ where $\nu$ and $z$
are the exponents associated with the quantum critical point and $d$ is the spatial dimension of the system.

\begin{figure}[htb]
\includegraphics[height=2.0in,width=3.1in, angle = 0]{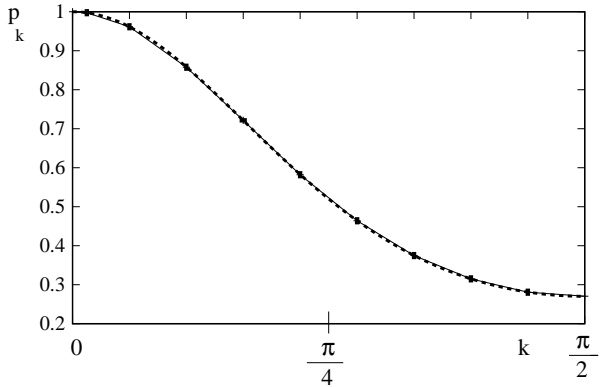}
\caption{$p_k$ vs k as obtained numerically (solid line) and analytically (dashed line) for $h_0 = 20$, $\omega = 0.0003$, $|J_x - J_y| = 0.05$, and $J = 1$.}
\end{figure}

\begin{figure}[htb]
\includegraphics[height=2.0in,width=3.1in, angle = 0]{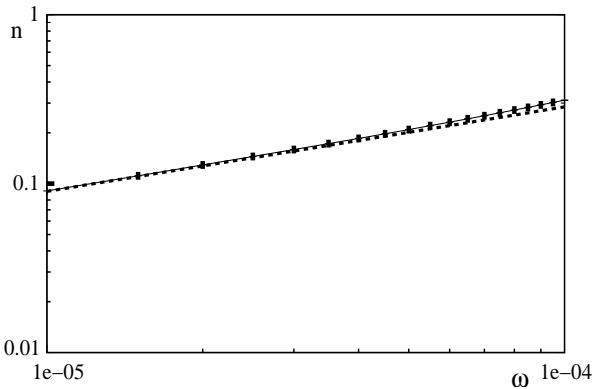}
\caption{n vs $\omega$ for $h_0 = 20$, $|J_x - J_y| = 0.05$ and $J = 1$.
 Solid line is found by numerically integrating $p_k$ over $k$, and 
the dashed line is the plot of eqn. (11).
Analytical and numerical results match exactly for lower values of $\omega$.} 
\end{figure}

\subsection{Entropy}
Following a recent paper by R. Barankov and A. polkovnikov\ct{polkovnikov08}, we define the  diagonal entropy $s_{d}(k)$ for each $k$ mode by 
\ba
s_{d}(k) = -\sum_l {\rho_{ll,k}ln\rho_{ll,k}}
\ea
where $\rho_{ll,k} = \bra l|\rho_k|l\ket$, $\rho_k$ being the instantaneous density matrix of the system for the mode $k$. One advantage of using the diagonal entropy is that, it follows the thermodynamical relations as expected to be followed by entropy defined at higher temperatures.  The diagonal entropy becomes identical to the previously defined Von Neumann entropy ($s_{VN}$), given by $s_{VN} = -\int^{\pi}_{0}{tr(\rho_k ln{\rho_k}) dk}/\pi$, when the off-diagonal terms in the density matrix, evaluated at the final time, go to zero upon coarse graining over $k$ space \ct{levitov06,mukherjee07,mukherjee08}. 
We have checked the variation of the diagonal entropy density, evaluated at the final time, with $\omega$ by numerically integrating $s_d(k) = p_k ln p_k + (1 - p_k) ln (1 - p_k) $ over all $k$, with $p_k$ obtained from eqn. (10). It is seen that the entropy attains a maximum near $\omega = \omega _0 = \pi (J_x - J_y)^2/h_0 ln2$ where the  non-adiabatic transition probability (see Eq.~10) for the mode $k=\pi/2$ is half, and falls off on either side of $\omega _0$. It should be noted that $\omega_0$ closely resembles the characteristic time scale $\tau_0$ appearing in the case of linear quenching \ct{levitov06}.

\begin{figure}[htb]
\includegraphics[height=2.0in,width=3.1in, angle = 0]{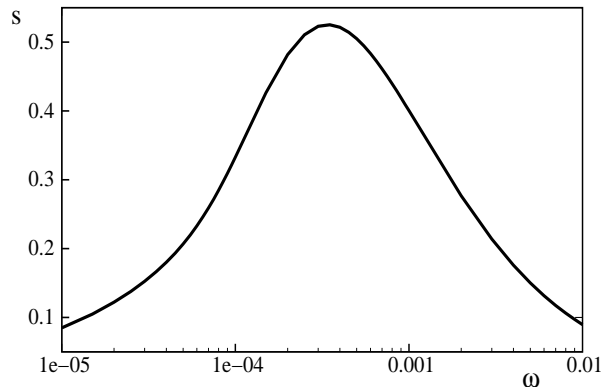}
\caption{Variation of diagonal entropy density with $\omega$ for one half cycle with $h_0 = 20, |J_x - J_y| = 0.05$ and $J = 1$,
 as obtained by numerical integration of $s_d(k)$ using the analytical expressions of $p_k$ (eq. 10).
 The entropy for one half cycle attains maxima near $\omega \sim \omega _0 = \pi (J_x - J_y)^2/h_0 \ln 2$.} 
\end{figure}

\section{Oscillatory quenching through a quantum critical point: multiple crossings}

Let us now focus on the repeated quenching case when the spin chain is periodically driven
through the quantum critical point. In the present section, it will be shown that interference plays a major role in deciding the behaviour of the system, and for some choices of parameters, the phases will add up destructively to make the tunneling probability almost zero. In order to be able to treat the successive Landau Zener transitions ,  as independent events, the time between two successive crossings has to be greater than the Landau Zener transition time for a single crossing as mentioned in the previous section.
To attain this limit we shall once again restrict our study to large values of $h_0$
and small $\omega$.  The system is prepared in the state $|1\ket$ at time $t = 0$. The diagonal terms of the Hamiltonian for each modes $k$ vanish, and consequently the gap becomes minimum, when the magnetic field crosses the $-J\cos k$ line as shown in Fig. ~4.

When the system approaches the crossing points of the diabatic levels, the energy gap is minimum leading to large relaxation time  and 
the system  fails to evolve adiabatically and the non-adiabatic transitions take place. 0n the other hand, away from the crossing points, the system follows adiabatic dynamics. Consequently, the evolution matrices associated with the system are different for close to and away from the crossing points \ct{hanggi05,ashhab07}.
\begin{figure}[htb]
\includegraphics[height=2.0in,width=3.1in, angle = 0]{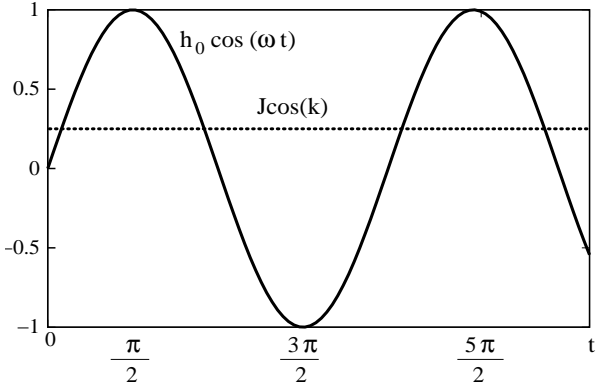}
\caption{Variation of energy levels due to the application of an oscillatory transverse field. The gap becomes minimum at the points where the magnetic field (shown by oscillatory solid line) becomes equal to $-J \cos k$.} 
\end{figure}
Between the crossings, the system evolves following the matrix
\ba
 G_j = && \left[ \begin{array}{cc} e^{-i\theta_j} & 0 \\
0 & e^{i\theta_j} \end{array} \right], \non \ea
where $j$ denotes the direction in which the system goes through  the crossing points.
 The LZ crossing in the non-adiabatic region can be approximately described by the evolution matrix
\ba
G_{LZ,j} = && \left[ \begin{array}{cc} \cos{\frac{\chi}{2}} & \sin{\frac{\chi}{2}}e^{i\theta_{LZ,j}} \\
-\sin{\frac{\chi}{2}}e^{-i\theta_{LZ,j}} & \cos{\frac{\chi}{2}} \end{array} \right], \non \ea
where the angle $\chi$ is given by 
\be \sin^2 {\frac{\chi}{2}} = 1 - \exp(-2\pi\gamma) \ee
where $\gamma$ is defined in the previous section and 
$j$ once again defines the direction of quenching (with respect to the crossing point). We have suppressed the notation $k$ denoting the wave vector for the time being. Also
\be
\theta_{LZ,1} \approx \frac{\pi}{2} + \theta_{stokes} \ee \
\be \theta_{LZ,2} \approx \frac{\pi}{2} - \theta_{stokes} \ee \
\be \theta_{stokes} = \frac{\pi}{4} + arg\Gamma(1 - i\gamma) + \gamma(ln\gamma - 1)
\ee \
where $\Gamma (x) $ is the gamma function and $\theta_{stokes} \rightarrow \pi/4$ or $\theta_{stokes} \rightarrow 0$, as $\gamma \rightarrow 0$ or $\gamma \rightarrow \infty$ respectively.  If the system is repeatedly quenched with the sinusoidal field a series of Landau Zener crossings take place with the evolution of the system described by the successive application of matrices defined above. More specifically for one full cycle, i.e., $\omega t$ going from $0$ to $2\pi$, one can write the complete evolution matrix as a product of $G_j$ and $G_{LZ,j}$, given by
\ba
G = G_{LZ,2} G_2 G_{LZ,1} G_1 , \ea
which can be generalized  for $N$ complete cycles to the form
\ba
G_N = (G_{LZ,2} G_2 G_{LZ,1} G_1)^N~. \ea
The probability amplitude of the states $ |i\ket$ $C_{i,N}$ ($i = 1,2$) at the final time  $\omega t = 2N\pi$ therefore obeys the
relation
\ba
\left[\begin{array}{cc} C_{1,N} \\
C_{2,N} \end{array} \right]     = (G_{LZ,2} G_2 G_{LZ,1} G_1)^N \left[\begin{array}{cc} C_{1,0} \\
C_{2,0} \end{array} \right]  
\ea
 where $C_{1,0}=1$ and $C_{2,0}=0$ at initial time $t=0$.
A little bit of algebra yields
\ba
 G_{LZ,2} G_2 G_{LZ,1} G_1 = && \left[\begin{array}{cc} g_{11} & g_{21} \\
                                       -g^*_{21} & g^*_{11}
                                      \end{array} \right]
\ea
where
\ba
&& g_{11} = \cos^2{\frac{\chi}{2}} e^{-i(\theta_1 + \theta_2)} - \sin^2{\frac{\chi}{2}} e^{i(-\theta_{LZ,1} + \theta_{LZ,2} - \theta_1 + \theta_2)}  \non \\
 && g_{21} = \sin{\frac{\chi}{2}}\cos{\frac{\chi}{2}} (e^{i(\theta_{LZ,1} + \theta_1 - \theta_2)} + e^{i(\theta_{LZ,2 + \theta_1 + \theta_2})})\ea
Denoting the probability that for mode $k$ the system is in state $ |2\ket$ after the $n$th crossing by $Q_{n,k}$, we get,
\ba
Q_{1,k} = (1 - p_k) 
\ea
as seen in the previous section and for one complete full period of oscillation
\ba
Q_{2,k} = 4p_k (1 - p_k) \sin ^2 {(\theta_{stokes} + \theta_2)}
\ea
For small anisotropy i.e., $(J_x - J_y)^2 << \omega\sqrt{h^2_0 - J^2 \cos^2 k}$ we have $\theta_{stokes} \rightarrow \pi/4$, and $\theta_1$ and $\theta_2$ are given by
\ba
\theta_2 &=& 2\int ^{\pi /2\omega }_0 {\sqrt {(h_0 \cos {\omega t} + J\cos k)^2 + \Delta_k^2}} \non \\ && \approx (\frac{2h_0 + J\pi \cos k}{\omega })
\ea
\ba
\theta_1 &=& 2\int ^{\pi}_{\pi /2\omega } {\sqrt {(h_0 \cos {\omega t} + J\cos k)^2 +\Delta_k^2 }} \non \\ && \approx (\frac{-2h_0 + J\pi \cos k}{\omega }). 
\ea
Subsituting Eqs.~24 and 25 in Eq.~23, 
we get\ct{kayanuma92,kayanuma94}
\ba
Q_{2,k} = 4p_k (1 - p_k) \sin ^2 {(\frac{2h_0 + J\pi \cos k}{\omega } + \frac {\pi }{4})}. \ea
$Q_{2,k}$ as obtained numerically and analytically as given in Eq.~26 are plotted as a function of $k$ in fig. (5) and (6). The numerical plot is fairly in agreement with the analytical results. It is seen that $Q_{2,k}$ oscillates, with the tunneling probability going to zero for many $k$'s, showing the signatures of constructive and destructive interferences.

It is clear that  for very small $\omega$, $Q_{2,k}$ oscillates rapidly with $k$ due to the presence of the sinusoidal term in eq. (26). As a result the coarse grained or average $Q_{2,k}$ (denoted by $\overline Q_{2,k}$), obtained by integrating each $Q_{2,k}$ over a small range around that $k$ followed by normalization, gives 
\ba
\overline Q_{2,k} = 2p_k(1 - p_k),
\ea
 as obtained previously for repetition under linear quenching\ct{mukherjee08}. It has been shown in recent works \ct{mukherjee07, mukherjee08, levitov06} that, for linear quenching, we can evaluate the transition probabilities at the end of each cycle by using the coarse grained density matrix only, thereby simplifying the problem greatly by neglecting the off-diagonal terms in the matrix. Analogously, in the present case also, the characterstic time scale associated with the rate of change of the off-diagonal terms in the density matrix for a mode $k$ sets the critical value of $\omega$ below which we can safely denote the tunneling probability by the coarse grained expression of $Q_{2k}$ only, thereby yielding eq. (27)\ct{kayanuma92}. One concludes  that in the limit of  very small $\omega$, the time interval between two successive crossings is large enough to destroy the phase information in the coarse grained probabilities.

\begin{figure}[htb]
\includegraphics[height=2.0in,width=3.1in, angle = 0]{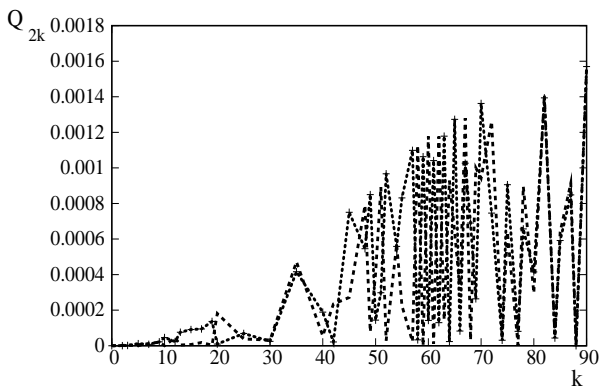}
\caption{Variation of $Q_{2,k}$ with $k$ for $h_0 = 20, |J_x - J_y| = 0.005$, $\omega = 0.01$ and $J = 1$. 
The widely spaced dashed line is analytical, and the numerical data points are shown on the 
closely spaced dashed line.} 
\end{figure}

\begin{figure}[htb]
\includegraphics[height=2.0in,width=3.1in, angle = 0]{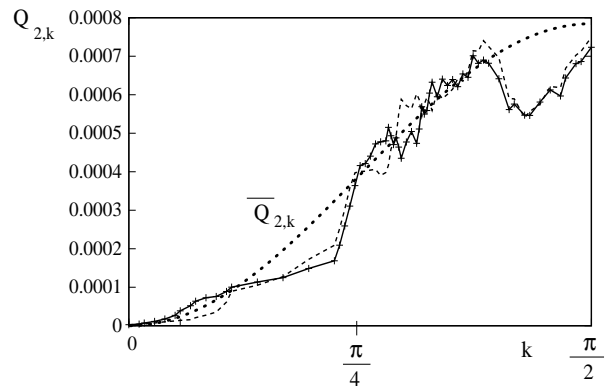}
\caption{Variation of $Q_{2,k}$ with $k$ for $h_0 = 20, |J_x - J_y| = 0.005$, $\omega = 0.01$ 
and $J = 1$ obtained by averaging out the oscillations of the data shown in fig. 5. 
The dashed line is analytical, and the solid line is numerical. The smooth dotted line is the 
plot of course grained excitation probability $\overline{Q_{2,k}}$ as obtained from eq. (27). } 
\end{figure}

\begin{figure}[htb]
\includegraphics[height=2.0in,width=3.1in, angle = 0]{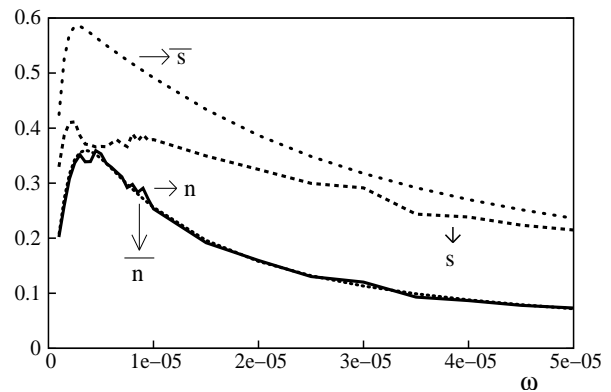}
\caption{Variation of actual and course grained kink density and entropy density with $\omega$
 for $h_0 = 20, |J_x - J_y| = 0.005$ and $J = 1$ for one complete cycle, obtained by numerically 
integrating $Q_{2,k}$ and $\overline {Q_2k}$ in eq. (26) and eq. (27) respectively. 
The actual and course grained plots for kink density match reasonably well, whereas we see a significant 
difference in case of entropy density, eventhough their qualitative behaviours show similarity 
for a wide range of $\omega$. The peaks of the plots occur near $\omega_0$.} 
\end{figure}

To generalize to the case of many complete periods,  it is useful to recast  eqn. (20) in the form  as\ct{ashhab07}
\ba
 G_{LZ,2} G_2 G_{LZ,1} G_1 = && \left(\begin{array}{cc} \cos {\frac{\zeta}{2}} & \sin{\frac{\zeta}{2}e^{i\phi}} \\
                                       -\sin{\frac{\zeta}{2}e^{-i\phi}} & \cos {\frac{\zeta}{2}}
                                      \end{array} \right) \non \\
&& \left(\begin{array}{cc} e^{-i\theta/2} & 0 \\
                                       0 & e^{i\theta/2}
                                      \end{array} \right)
\ea
We shall call the diagonal matrix as $U_1$ and the other as $U_2$ which successively operate on the column 
matrix $(C_{1,k}, C_{2.k})$.
 The angles $\zeta$ and $\theta$ are given as
\ba
\sin^2 {\frac{\zeta}{2}} \approx 4\sin^2{\frac{\chi}{2}}\cos^2(\frac{\theta_{LZ,1} - \theta_{LZ,2}}{2} - \theta_2) \ea 

\ba \theta = \tan^{-1}{\frac{A}{B}} \ea
where
\ba A &=& \cos^2{\frac{\chi}{2}}\sin{(\theta_1 + \theta_2)} \non \\ && + \sin^2{\frac{\chi}{2}}\sin{(\theta_{LZ,1} - \theta_{LZ,2} + \theta_1 - \theta_2)} \non \ea
and 
\ba B &=& \cos^2{\frac{\chi}{2}}\cos{(\theta_1 + \theta_2)} \non \\ && - \sin^2{\frac{\chi}{2}}\cos{(\theta_{LZ,1} - \theta_{LZ,2} + \theta_1 - \theta_2)} \non \ea

\be
\phi \approx \frac{\theta_{LZ,1} + \theta_{LZ,2}}{2} - \theta_2 \ee

The dynamics described by Eq.~(28) can be understood by exploring the properties of the rotation matrices  $U_1$ and $U_2$.  The role of $U_2$ is to rotate a vector about an axis in the $x-y$ plane by an angle of $\zeta$, whereas the  matrix $U_1$ brings about a rotation of the vector by an angle $\theta$ around the $z$ axis only \ct{goldstein}. If $\theta$ is a multiple of $2\pi$, which we call the resonance condition, the $z$-axis
rotation does not affect the dynamics and the small oscillations of $\zeta$ add up  constructively to produce full oscillations between
the diabatic states $|1\ket$ and $|2\ket$. On the other hand, if $\theta$ differs from a multiple of $2\pi$ by more than $\zeta$, then the rotations about angle $\zeta$ do not add up constructively, and the oscillations will be suppressed, thus resulting in an effective rotation about an axis almost parallel to the $z$ axis only \ct{ashhab07} (see figure 9).  In the present context, the resonance condition is given by,
\ba
\theta \approx 2(\theta_1 + \theta_2) = \frac{4\pi J\cos k}{\omega} = 2n\pi, \ea  i.e.,
\ba
\frac{2 J\cos k}{\omega} = n \ea
Therefore for the resonance conditions, we can write
\ba
\left[\begin{array}{cc} C_{1,N} \\
C_{2,N} \end{array} \right] \non    = \pm \left(\begin{array}{cc} \cos {\frac{\zeta}{2}} & \sin{\frac{\zeta}{2}e^{i\phi}} \\
                                       -\sin{\frac{\zeta}{2}e^{-i\phi}} & \cos {\frac{\zeta}{2}}
                                      \end{array} \right)^N \left[\begin{array}{cc} C_{1,0} \\
C_{2,0} \end{array} \right] \non
\ea

\begin{figure}[htb]
\includegraphics[height=2.0in,width=3.1in, angle = 0]{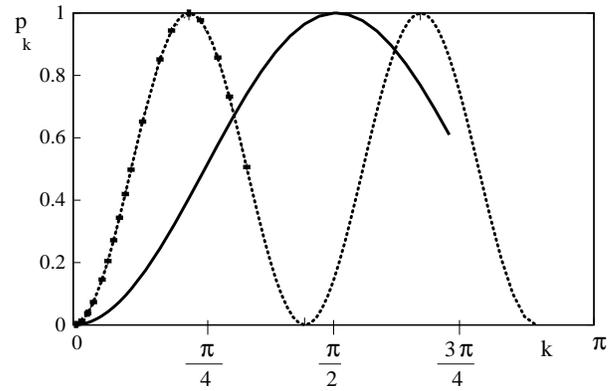}
\caption{Graph showing the behaviour of the excitation probability as a function of number of full cycles for $h_0 = 20, (J_x - J_y) = 0.005$, $J = 1$ and $\omega = 0.01$. The dashed line is obtained analytically for $k = 81.9521^0$ with integral $2J\cos k /\omega = 28$, while the solid line is the analytical graph for $k = 25.8419^0$ with integral $2J\cos k /\omega = 180$. Numerical data points shown on the dashed line corresponding to $k = 81.9521^0$ match exactly with the analytical results.} 
\end{figure}

\begin{figure}[htb]
\includegraphics[height=2.0in,width=3.1in, angle = 0]{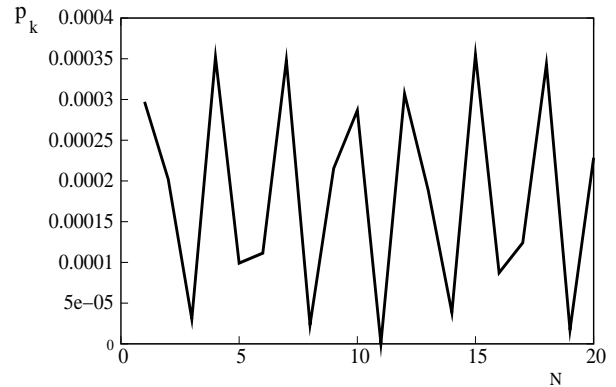}
\caption{Graph showing the behaviour of the excitation probability as a function of number of 
full cycles, obtained analytically for $h_0 = 20, |J_x - J_y| = 0.005$, $J = 1$, $\omega = 0.01, k = 80^0$ 
and non-integral $2J\cos k /\omega$. As expected, the excitation probability varies randomly 
and does not differ appreciably from its initial value.} 
\end{figure}

\begin{figure}[htb]
\includegraphics[height=2.0in,width=3.1in, angle = 0]{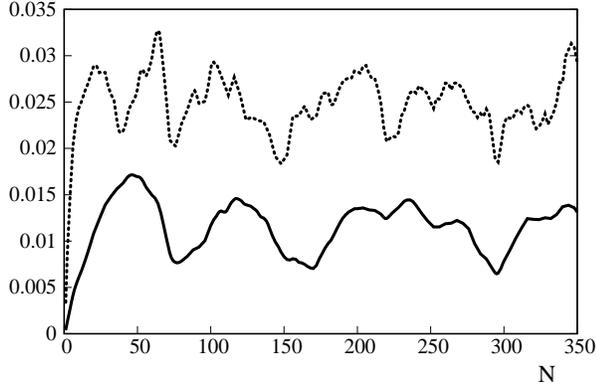}
\caption{Graph showing the behaviour of the kink density (solid line) and entropy density (dashed line) as a function 
of number of full cycles for $h_0 = 20, |J_x - J_y| = 0.005$, $J = 1$, and $\omega = 0.01$, 
as obtained by numerically integrating the excitation probabilities obtained by using eqn. (19). } 
\end{figure}

From this formalism it is clearly seen that for the resonance conditions, after $N$ complete cycles, we get $N$ successive rotations by the small angle $\zeta$. This causes oscillations in the probabilities of the two states with frequency given by
\ba\Omega &=& \frac{\zeta}{2\pi/\omega} = \frac{\omega \zeta}{2\pi} \non \\ &=& \frac{\omega}{\pi}\sin^{-1}[2\sqrt{1 - p_k} \non \\ && \cos{(\theta_{stokes} - \frac{2h_0 + J\pi\cos k}{\omega})}].\ea

Since  $\zeta$ depends on the wave vector $k$, the probabilities for each resonant  mode oscillates with its own characteristic frequency. As a result the kink density as well as the entropy density obtained by integrating over all modes shows an oscillatory behaviour (see figure 10). The oscillatory behaviour of the entropy density  observed here is an artifact of retaining the phase information  of the off-diagonal terms of the density matrix. It can be shown that in absence of phase information, $|C_{1,k}(\overline t)|^2, |C_{2,k}(\overline t)|^2 \rightarrow 1/2$ after each crossing, and as a result the entropy density of the system increases monotonically \ct{mukherjee08} with the number of crossings.

\section{quenching by a magnetic field varying both linearly and periodically }

In this section, we study the defect generation in a transverse XY spin chain driven by
a time-dependent magnetic field $h(t)$ which consists of a linear part as well as
an oscillatory part given by $h(t)= {t}/{\tau} + h_0 \cos\omega t$ where $\tau$ denotes
the rate of the linear part of the quenching. In the limit $h_0 \to 0$, the dynamics
reduces to the well studied Kibble-Zurek problem of linear quenching while in the other
limit of $\tau \to \infty$, one should recover the results presented in earlier sections.
The presence of both the linear and periodic terms non-trivially modifies the density of
defect in the final state as shown below. The reduced Hamiltonian in the present situation is given
by\ba &&H_k (t) = \non \\  && \left[ \begin{array}{cc} \epsilon & i (J_x - J_y)\sin k \\
-i(J_x - J_y) \sin k & -\epsilon \end{array} \right], \non \ea
with $\epsilon = t/\tau + h_0 \cos {\omega t} +J\cos k$ and we shall once again
recall the parametes $\alpha_k$ and $\Delta_k$ as defined before.

\begin{figure}[htb]
\includegraphics[height=2.0in,width=3.1in, angle = 0]{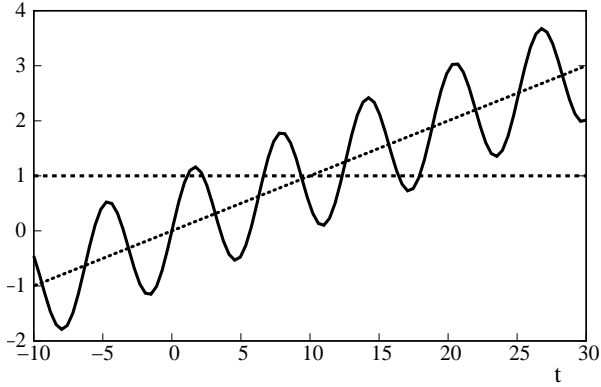}
\caption{Graph showing the behaviour of the diabatic energy levels with time when a linearly varying magnetic field is applied in addition to the oscillatory term.
 The inclined dotted line is the plot of $t/10$, the solid line is $\frac{t}{10} + \cos {t}$, and the dashed line parallel to $x$ axis is the constant $-J\cos k = 1$.} 
\end{figure}
For a given wavevector $k$, the instantaneous energy gap is minimum 
 at times $t_{0,k}$ such that
\ba
\frac{t_{0,k}}{\tau} + h_0 \cos {\omega t_{0,k}} + J\cos k = 0
\ea
Since $\cos {\omega t_{0,k}} \le 1$ always, so from the above equation we can conclude that all the $t_{0,k}$'s occur in the time interval $-\tau (J + h_0) < t_{0,k} < \tau (J + h_0)$. Also, since time between two successive $t_{0,k}$'s is of the order of $\pi /\omega$, so we can estimate the number of times that the gap goes to minimum for any $k$ is $\sim \frac{2h_0 \tau}{\pi /\omega}$, with the minimum number of times being $1$. The situation is depicted in Fig.~11.

In the adiabatic limit ($\Delta_k^2/ \alpha_k >>1$) the Landau Zener transition time \ct{vitanov99,mullen89} ($\tau _{LZ}$) is given as $\tau _{LZ} \sim \Delta_k^2/\alpha_k$, and in the non-adiabatic case, we have $\tau _{LZ} \sim 1/\sqrt {\alpha_k}$. It
should be noted in the present case, the rate of change of diagonal terms $\alpha_k = |\frac{d}{dt}2(\frac{t}{\tau} + h_0 \cos {\omega t} + J\cos k)|_{t_{0,k}} = 2|\frac{1}{\tau} - h_0 \omega \sin {\omega t_{0,k}}|$. Hence for our theory to be valid, i.e., to get widely separated non-overlapping LZ transitions, we need $\tau << \frac{\pi}{\omega (J_x - J_y)}$ for the adiabatic  and $\sqrt {\tau} << \frac{\pi}{\omega}$ for the non-adiabatic situations.

We prepare the system in its ground state at $t\rightarrow-\infty$ with $|C_{1,k}(-\infty)|^2=1$ and the
probability of defect for the mode $k$ in the final state at $t\rightarrow +\infty$ is given by the
probability $|C_{1,k}(+\infty)|^2$. For the linear as well as periodic driving, Eq.~8
when linearized around the crossing point $t = t_{0,k}$ gets modified to

\ba
\frac{d^2 C_{1,k}}{d^2 t} &+& 2i((\frac 1 {\tau} - h_0 \omega \sin \omega t_{0,k})(t - t_{0,k}))\frac {d C_{1,k}}{dt} \non \\ &+&
   \Delta_k^2 C_{1,k} = 0,
\ea
which leads to the non-adiabatic transition probability
\ba
p_{k} = e^{-\frac{\pi \Delta_k^2}{\frac{1}{\tau} - h_0 \omega \sin {\omega t_{0,k}}}}
\ea
 
In the limit of small $\tau$ and large $\omega$, we can expand the excitation probability as $p_k \approx 1 - \frac{\pi (J_x - J_y)^2 \sin^2 {k}}{1/\tau - h_0\omega \sin {\omega t_{0,k}}}$. Further, for $\frac{1}{\tau} >> h_0\omega\sin{\omega t_{0,k}}$, we can write the expression for kink density as 
\ba
n &\approx& \frac{1}{\pi}\int^{\pi}_{0}{[1 - \frac{\pi (J_x - J_y)^2 \sin^2 {k}}{1/\tau - h_0\omega \sin {\omega t_{0,k}}}] dk} \non \\ &\approx& 1 - \frac{\pi (J_x - J_y)^2}{2}\tau . 
\ea
The approximate equation given in Eq.~(38) matches perfectly with the numerical integration results (see Fig.~13).
On the other hand, in the limit of large $\tau$ and small $\omega$, only the modes close to $k = 0$ or $\pi$  contribute to the defect density, and by considering only the $0$ and $k$ modes in $t_{0,k}$, we can arrive at an approximate 
analytical expression given by 
\ba
n = \frac{1}{\pi}\int^{\pi}_{0}{p_k dk} &=& \frac{\pi \sqrt {\frac{1}{\tau} - h_0 \omega \sin {\omega t_{0,0}}}}{2|J_x - J_y|} \non \\ &+& \frac{\pi \sqrt {\frac{1}{\tau} - h_0 \omega \sin {\omega t_{0,\pi}}}}{2|J_x - J_y|}
\ea
The kink density as a function of $\tau$ for the non-adiabatic and adiabatic cases, as obtained from equations (38) and (39) respectively, are plotted in Figs.~13 and 14 together with the corresponding numerically obtained values. As expected, in case of non-adiabatic evolution, we get exact matching between analytical and numerical results only for low values of $\tau$ for which $\frac{1}{\tau} >> h_0 \omega \sin{t_{0,k}}$, whereas in case of adiabatic evolution, we see good agreement between the analytical and numerical results only when $\frac{1}{\tau}$ is not close to $h_0\sin{\omega t_{0,0}}$ or $h_0\sin{\omega t_{0,\pi}}$, since around these values of $\tau$, the effects of $h_0\sin{\omega t_{0,k}}$ for $k \ne 0,\pi$ become important.

It should be noted that for the single crossing case, using the similar line of arguments
given in section II,  one can propose a generalized Kibble- Zurek scaling form of the defect 
density in the final state given as 
\ba
n &\sim& a_0[|\frac 1{\tau} - h_0 \omega \sin\omega t_{0,0}|^{\nu d/(\nu z+1)}] \non \\ &+& a_{\pi}[|\frac 1{\tau} - h_0 \omega \sin\omega t_{0,\pi}|^{\nu d/(\nu z+1)}],
\ea
where $a_0$ and $a_{\pi}$ are two constants.

\begin{figure}[htb]
\includegraphics[height=2.0in,width=3.1in, angle = 0]{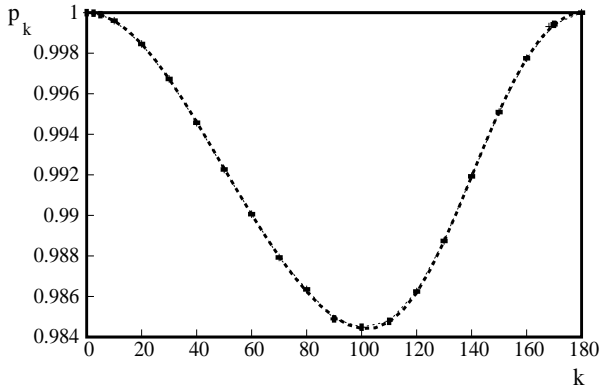}
\caption{$p_k$ vs $k$ (in degrees) for the case when gap becomes minimum only once, with $\tau = 2,h_0 = 1, \omega = 0.1, |J_x - J_y| = 0.05$ and $J = 10$. 
The dashed line is analytical and the numerical results shown as data points coinside 
exactly with the analytical values.} 
\end{figure}

\begin{figure}[htb]
\includegraphics[height=2.0in,width=3.1in, angle = 0]{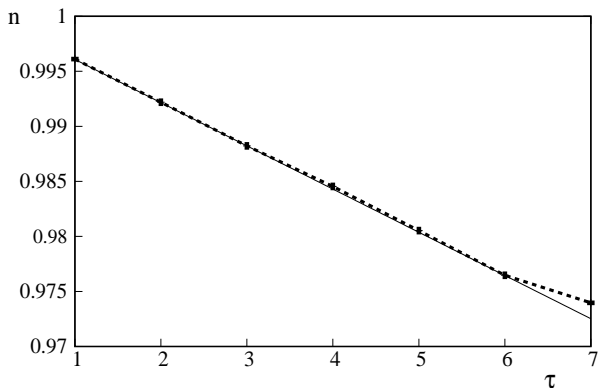}
\caption{Kink density ($n$) vs $\tau$ for $h_0 = 1, \omega = 0.1, |J_x - J_y| = 0.05$ and $J = 10$. The solid line is the plot of eq. (38), and the dashed line is obtained by numerical integration of the analytical expression of $p_k$ as given in eq. (37). We get exact matching between the two results for low $\tau$ only, as expected from the theory. Only single crossing occurs for the range of $\tau$ shown in the figure.} 
\end{figure}

\begin{figure}[htb]
\includegraphics[height=2.0in,width=3.1in, angle = 0]{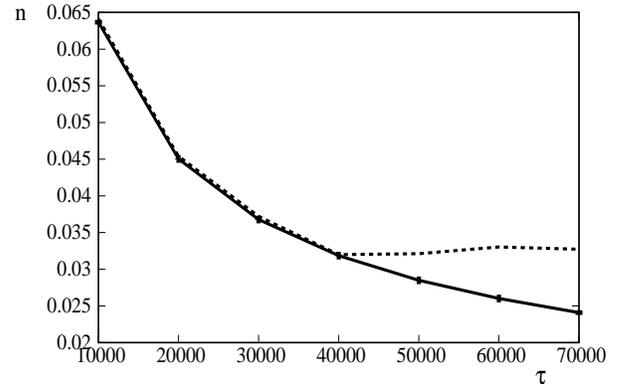}
\caption{Kink density $n$ vs $\tau$ for $h_0 = 1, \omega = 0.000001, \Delta = 0.05$ and $J = 10$. Only single crossing occurs for the range of $\tau$ shown in the figure. The dashed line is found by numerical integration of eq. (37), and the solid line is obtained by using eq. (39). We get exact matching between the two results except when $\frac{1}{\tau}$ is very close to $h_0\omega\sin{\omega t_{0,0}}$ or $h_0\omega\sin{\omega t_{0,\pi}}$.} 
\end{figure}

Now we concentrate on the situation of multiple crossings of the energy minima. For multiple crossings to occur for a given wave vector  $k$, Fig. ~11  implies that there should  exist  a $t = \overline t$ such that $\frac{1}{\tau} - h_0 \omega \sin {\omega \overline t } = 0$, i.e., $\sin {\omega \overline t } = \frac{1}{\tau h_0 \omega}$.  This  is possible only if $|\frac{1}{\tau h_0 \omega}| \le 1$. The Schrodinger equations for the probability amplitudes can be put in the form

\ba
\frac{dC_{1,k}}{dt} = \Delta_k e^{i(\frac{t^2}{\tau} + \frac{2h_0 \sin {\omega t}}{\omega } + 2Jt\cos k)}C_{2,k}\ea

\ba 
\frac{dC_{2,k}}{dt} = \Delta_k e^{i(\frac{t^2}{\tau} + \frac{2h_0 \sin {\omega t}}{\omega } + 2Jt\cos k)}C_{1,k}\ea

Using the relation $e^{\pm iz\sin {\omega t}} = \sum ^{\infty}_{r = -\infty }J_r(z)e^{\pm ir\omega t}$, where $J_r(z)$ is the Bessel function of first kind of order $r$, given by
\ba
J_r(\eta) = \Sigma ^{\infty}_{m=0}\frac{(-1)^m}{m!\Gamma(m+r+1)}(\frac{\eta}{2})^{2m+1},
\ea
we recast the equation to the form
\ba
\frac{dC_{1,k}}{dt} = \Delta_k \sum^{\infty}_{r = -\infty } {J_r (\frac{2h_0}{\omega}) e^{i(\frac{t^2}{\tau} + 2Jt\cos k + r\omega t)}C_{2,k}}. \ea

The terms on the R.H.S. of eqn. (44) being  rapidly varying in time,  $\frac{dC_{1,k}}{dt}$ attains a non-negligible value only when the phase is stationary. Using 
\ba
&& \frac{t^2}{\tau} + 2Jt\cos k + r\omega t \non \\ &=& \frac{1}{\tau}(t + \frac{2J\tau\cos k + r\omega \tau}{2})^2 - \frac{(2J\tau \cos k + r\omega \tau)^2}{4\tau} \ea
we find that  $\frac{dC_{1,k}}{dt}$ is non-negligible only close to  $t = -\frac{2J\tau \cos k + r\omega \tau}{2}$, with $r = 0, \pm 1, \pm 2, ...$.  The above relation implies  the
existence of an effective phase transition for each value of $r$.  Choosing $r = l$ (say)  and denoting $C_{i,k}$ by $C_{i,k,l}$ $(i = 1,2)$,  we get
\ba
\frac{dC_{1,l,k}}{dt} &=& \Delta_k J_l e^{-i\frac{(2J\tau \cos k + l\omega \tau)^2}{4\tau }}\non \\ && e^{i\frac{1}{\tau}[t + \frac{2J\tau \cos {k} + l\omega \tau}{2}]^2} C_{2,l,k} \ea

Invoking upon the transformation to a new variable $s = t + l\omega \tau /2$, we get
\ba
\frac{dC_{1,l,k}}{ds} &=& \Delta_k J_l e^{-i \frac{l^2 \omega ^2 \tau ^2 + 4J\tau ^2 l \omega \cos k}{4\tau}} \non \\ && e^{-iJ^2 \tau \cos ^2 k}e^{\frac{i}{\tau}(s + J\tau \cos k)^2} C_{2,l,k}\ea
Let us compare with purely linearly quenching case ($h_0=0)$, when the above equation gets modified to
\ba
\frac{dC_{1,l,k}}{ds} = \Delta_k e^{-i J^2 \tau \cos ^2 k}e^{\frac{i}{\tau}(s + J\tau \cos k)^2}C_{2,l,k} \ea
The role of periodic modulation on top of the linear quenching is to  renormalize   $\Delta _k$
to $\overline{\Delta_k}$  with $\overline{\Delta_k}=\Delta _k J_l e^{-i\frac{l^2 \omega ^2 \tau ^2 + 4J \tau ^2 l \omega \cos k}{4\tau}} $. Note that $\overline{\Delta_k}$ also 
vanishes at the quantum critical point for the modes $k=0$ and $\pi$.

The dividing of the probability amplitudes for a given wave vector to different $l$ values by using  the Bessel's function can be visualized in the following way:  The two energy levels for the wave vector $k$ are
assumed to consist of a number of sublevels\ct{kayanuma00}, with probability amplitudes of the $l$th level being denoted by $C_{1,l,k}$ and $C_{2,l.k}$. Each sublevel undergoes a level crossing only once through the course of dynamics, and for the $l$th transition for the  mode $k$, the incoming state (given by $\overrightarrow{C_{l-1,k}}$) and the outgoing state (given by $\overrightarrow{C_{l,k}}$) are connected by the transfer matrix \ct{kayanuma00}
\ba &M_{l,k}& \non \\ &=& \left[ \begin{array}{cc} D _{l,k} & \beta _{l,k} e^{-i\frac{l^2 \omega ^2 \tau ^2 + 4J\tau ^2 l\omega \cos k}{4\tau}} \\
-\beta^* _{l,k} e^{-i\frac{l^2 \omega ^2 \tau ^2 + 4J\tau ^2 l\omega \cos k}{4\tau}} & D_{l,k} \end{array} \right], \non \ea
where $D_{l,k} = \sqrt{p_{l,k}}$ and $\beta_{l,k} = sgn J_l(\eta) \sqrt {1 - p_{l,k}} e^{-i\phi _{l,k}}$, in which $p_{l,k} = e^{-\pi \tau \Delta^2 (J_l(\eta))^2} $, $\eta = 2h_0/\omega$ and $\phi_{l,k}$ is the Stokes phase given by
\ba
\phi_{l,k} = \pi/4 + arg\Gamma(1 - i\delta_{l,k}) + \delta_{l,k}(ln\delta_{l,k} - 1)
\ea
where $\delta_{l,k} = \tau \Delta^2_k (J_l(\eta))^2 /2$, in terms of the gamma function $\Gamma(z)$.

It should be noted that $J_l(\eta)\rightarrow 0$ for large $l$, and $\Sigma^{\infty}_{l\rightarrow -\infty} J^2_l(\eta) = 1$. So the transition is confined to a finite region, and the infinite series of recursive relation for $l \rightarrow \infty$ converges to a finite value.  In case of $l$ for which $J_l(\eta) \approx 0$, $M_{l,k}$ gets reduced to an identity matrix. Hence taking $J_l(\eta) = 0$ $\forall l > l_f$ we can write the state vector 
 $\stackrel{\rightarrow}{C}_{l,k} = (C_{1,l,k}, C_{2,l,k})$ as 
\ba
\stackrel{\rightarrow}{C_{l,k}} &=& M_{l,k}\stackrel{\rightarrow}{C}_{l-1,k} \non \\ &=& M_{l,k}M_{l-1,k}...M_{0,k}...M_{-l_f + 1,k}M_{-l_f,k}\stackrel{\rightarrow}{C}_{in,k} \ea
where $\stackrel{\rightarrow}{C}_{in,k}$ denotes the initial condition. The probability of excitation at infinite time is given by $p_k(\infty) = |C_{1,k}(\infty)|^2$. 
\ba
p_k(\infty) &=& |C_{1,k}(\infty)|^2 \non \\ &=& |\left[ \begin{array}{cc} 1 & 0 \end{array} \right]M_{l_f,k}M_{l_f - 1,k}... \non \\ && M_{0,k}...M_{-l_f + 1,k}M_{-l_f,k}\stackrel{\rightarrow}{C}_{in,k}|^2 \ea

We are interested in evaluating the defect density in the final state in the limit of
large $\tau$ and hence the off-diagonal terms in the matrix $M_{l,k}$ vanish upon
coarse-graining, i.e., upon integration over all $k$. This approximation leads to the final result for the
coarse grained non-adiabatic transition probability given by
\ba
|C_{1,l,k}|^2 = p_{l,k}|C_{1,l-1,k}|^2 + (1 - p_{l,k})|C_{2,l-1,k}|^2
\ea
after neglecting the cross terms. The effects of the critical points are manifested by $\overline {\Delta_k}$ vanishing at $k = 0, \pi$ and those modes do not evolve from
their initial state.

  Comparison between kink density obtained by numerical integration of Schr\"{o}dinger equation and by using the approximate analytical  eqns. (51) and (52) for different $\tau$ has been shown in figure (15).  We also plot the diagonal entropy density in figures. (16) and (17).  Although, the defect density in the final state decays with increasing $\tau$, the entropy density is found to increase  and ultimately saturates as $\tau$ increases. This is in sharp contrast to the cases of linear quenching \ct{mukherjee08,mukherjee07} where it attains a maxima at a characteristic scale $\tau = \tau_0$, and falls off on either side of $\tau_0$. This behaviour can be explained in the following way:  for very small $\tau$ the system fails to evolve appreciably and therefore remains very close to its
initial ordered state at the final time.  For larger values of $\tau$,  the probabilities change which introduces disorder in the final state leading to  higher value of entropy density . For very large $\tau$, the linear term becomes insignificant compared to the oscillatory term in the Hamiltonian. Consequently, only the oscillatory quenching term contributes to the dynamics of the system, keeping the entropy of the system constant for a fixed value of $\omega$. It is to be noted that the interaction between the levels  depend on the value of $J_l(2h_0/\omega)$ which saturates in the limit of large $\omega$ if $h_0$ is held fixed irrespective of the values of $l$ and $k$. As a result,  the kink density and also  entropy density  saturate in the limit of large $\omega$, as shown in fig.~(17)

\begin{figure}[htb]
\includegraphics[height=2.0in,width=3.1in, angle = 0]{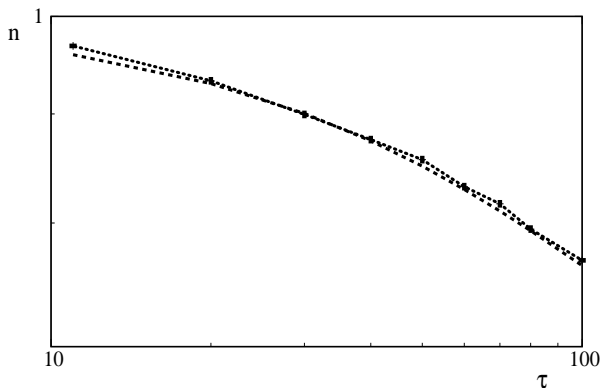}
\caption{Graph showing the variation of the kink density with $\tau$ for multiple crossings occurring with $h_0=1,~(J_x - J_y) = 0.05,~(J_x + J_y) = 10$, 
and $\omega = 0.1$ . Results obtained by numerically solving Schr\"{o}dinger equation (dotted line) match reasonably well with the analytical ones (dashed line).} 
\end{figure}

\begin{figure}[htb]
\includegraphics[height=2.0in,width=3.1in, angle = 0]{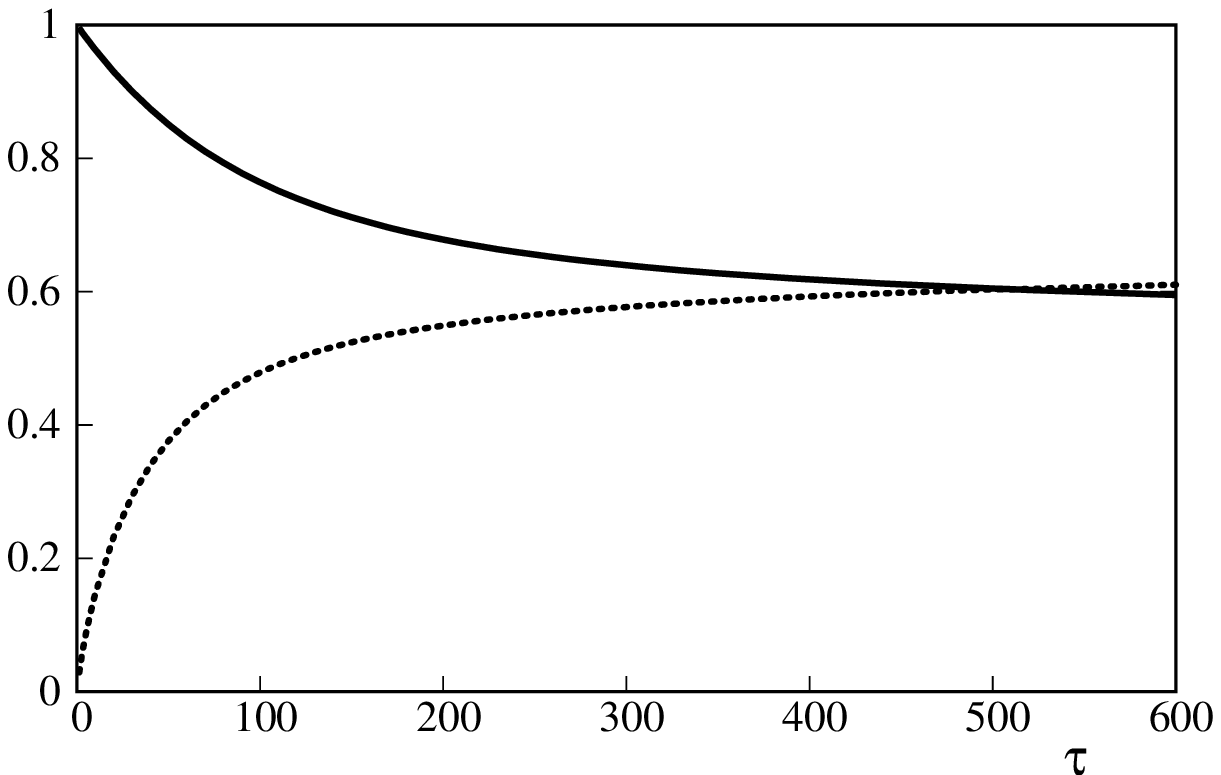}
\caption{Graph showing the variation of the kink density (solid line) and entropy density (dashed line)
 with $\tau$ as obtained analytically for $h_0=1,~(J_x - J_y) = 0.05,~(J_x + J_y) = 10$, 
and $\omega = 0.1$.} 
\end{figure}

\begin{figure}[htb]
\includegraphics[height= 2in,width = 3.1in, angle = 0]{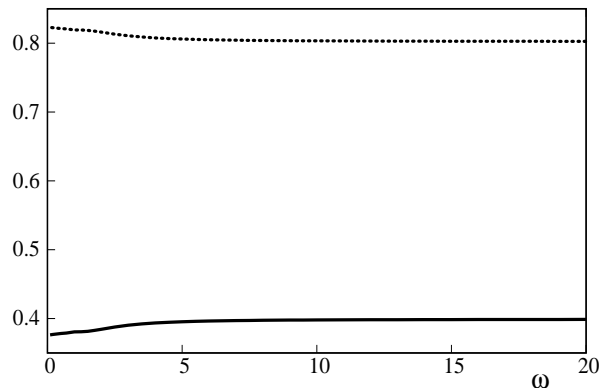}
\caption{Graph showing the variation of the kink density (solid line) and entropy density (dashed line)
 with $\omega$ as obtained analytically for $h_0=1,~(J_x - J_y) = 0.05,~(J_x + J_y) = 10$, 
and $\tau = 50$.} 
\end{figure}

\section{conclusion}

In conclusion, we have studied the effects of interference in the quenching dynamics of a one-dimensional transverse XY spin chains in the presence of a time-dependent magnetic field $h(t)=h_0 \cos \omega t$ or $h(t)=t/\tau + h_0 \cos \omega t$. The system is initially prepared in its ground state and we estimate the defect density and entropy density in the final state following the quench using both approximate analytical and direct numerical integration techniques. In all the situations, the analytical and numerical results are found to be in good
agreement. Our observations presented in the paper are summarized as follows.

Firstly, we have studied the defect density in the final state following a single crossing by linearizing the oscillatory magnetic field round the time at which  the instantaneous energy gap is minimmum. We show that in the limit of large $h_0$ and
small $\omega$, the defect density scales as $\sqrt{\omega}$. The observation is supported
by numerical solution of the Schrodinger equation in the limit of small $\omega$. On the
other hand, the diagonal entropy density shows a maximum at a characteristic frequency
scale $\omega_0$ as defined in the text. Effects of interference are invisible in the case of a single crossing only. We do also suggest an equivalent Kibble-Zurek
scaling relation for the defect density in the above situation.

In the next section we generalize to the multiple crossing situation where the interference
of the probability densities play a dominant role. We use two different transfer
matrices valid close to and away from the crossing points. We show that for a full cycle of
oscillation the results obtained for repeated linear quenching \ct{mukherjee08} when
the off-diagonal terms of the density matrix are coarse grained, leading to loss of phase information which gives rise to constructive and destructive interferences, is a valid 
approximation in the limit of small $\omega$. For multiple crossings, we show that
there exist resonance wave vectors for which the non-adiabatic transition probability oscillates
between zero and one with the number of crossings following a characteristic $k$-dependent
frequency. As a result the defect density also exhibits an oscillatory behaviour.  The entropy density also shows a similar dependence on the number of crossings, which is in stark contrast with the linear quenching case, in which exclusion of the interference effects in the excitation probabilities causes the entropy density to increase monotonically with the number of crossings. It may be noted that a similar oscillatory behaviour is observed
for the central spin of quantum Heisenberg chain \ct{yuan08}.

Lastly, we study the quenching of the spin chain in the presence of a magnetic field which
is varying linearly with time as $t/\tau$ and also modulated by a periodically varying
part $h_0\cos\omega t$. For the single crossing case, we once again use the linearization
method which predicts a defect density that is in fair agreement with the numerically
obtained result.For multiple crossings, we again invoke the transfer matrix approach to evaluate the cross-grained defect density. In this case it has been shown that for large values $\tau$ we can safely neglect the phase information, and hence the effects of interference, by coarse graining the density matrix. Our analytical and numerical results show that the defect density 
decreases with increasing $\tau$ for a given $\omega$ whereas when $\omega$ is varied
with $\tau$ fixed, the defect density saturates for higher values of $\omega$. The
entropy density also exhibits a monotonic increase as a function of $\tau$ with fixed
$\omega$, an observation that is in sharp contrast with the linear quenching case where
the entropy density attains a maximum at a characteristic time scale $\tau_0$ \ct{mukherjee07} . This may be an artifact of the integrability of the model which gets decoupled into independent two-level systems \ct{polkovnikov_private}.

\begin{center}
{\bf Acknowledgements}
\end{center}
AD acknowledges R. Moessener and the hospitality of MPIPKS, Dresden, where a major part of this work was done. We acknowledge Anatoli Polkovnikov, Diptiman Sen and Uma Divakaran
for critical comments and helpful discussions.

\end{document}